\documentclass{appolb}
\usepackage{graphicx}
\usepackage{amsmath}
\usepackage{rotating}
\usepackage{bm}      
\usepackage{amssymb}
\usepackage{amsmath}

\begin{document}
\title{SUPERHEAVY TOROIDAL NUCLEI IN THE SKYRME ENERGY FUNCTIONAL FRAMEWORK\thanks{Presented at the XXIII Nuclear Physics Workshop "Marie and Pierre Curie",\\
 Kazimierz Dolny, Poland, $27^{th}$ Sept. - $2^{nd}$ Oct., 2016}%
}
\author{A. Kosior$^{1}$, A. Staszczak$^{1}$ and Cheuk-Yin Wong$^{2}$
\address{$^{1}$Institute of Physics, Maria Curie-Sk{\l}odowska University, Lublin, Poland\\
$^{2}$Physics Division, Oak Ridge National Laboratory, Oak Ridge, TN USA}
}
\maketitle

\begin{abstract}
Using the Hartree-Fock-Bogoliubov (HFB) self-consistent mean-field \break
theory with the SkM* Skyrme energy-density functional, we study nuclear
structure properties of even-even superheavy nuclei (SHN) of $Z=120$ isotopes
and $N=184$  isotones. The shape of the nucleus along the lowest energy curve
as a function of the quadrupole moment $Q_{20}$ makes a sudden transition from
the oblate spheroids (biconcave discs) to the toroidal shapes, in the region of
large oblate quadrupole moments.
\end{abstract}

\section{Introduction}
Since the time when Wheeler coined the name \textit{superheavy nuclei} (SHN) in 1955 \cite{Whe55},
our knowledge of the ``very heavy nuclei" has become more extensive and systematic.
During the last 60 years the heaviest known nucleus limit has been extended from
$^{256}_{101}$Md (1955) \cite{Ghi55} to $^{294}_{118}$Og (2006) \cite{Oga06}, and
the properties of SHN have been studied mostly in the region of prolate deformations.
In this region, the energy surfaces of SHN reveal two paths to fission:
a reflection-symmetric path corresponding to elongated fission fragments (sEF)
and the reflection-asymmetric path with elongated fission fragments (aEF),
which bifurcates from the sEF path after the first barrier, see e.g.
Ref.~\cite{Sta13}. There are also predictions on the ground state deformations,
fission barrier heights, and spontaneous-fission and $\alpha$-decay half-lives
of SHN; for a recent review, see for example Refs.~\cite{Hee15,Bar15}.

Theoretically, the properties of SHN have been studied much less in the oblate
region than in the prolate region, with a few exceptions, such as the study on
super-deformed-oblate SHN at quadrupole moment $Q_{20}=-60$ to $-55$~b \cite{Jach11,Hee15}.
Within the self-consistent constraint Skyrme-Hartree-Fock+BCS model,
we found equilibrium toroidal nuclear density distributions at oblate
deformation $Q_{20}\leq -200$ b for the hypothetical SHN $^{316}122$,
$^{340}130$, $^{352}134$, and $^{364}138$ \cite{Sta08}.

It is interesting to note that it was also Wheeler who suggested long ago that
under appropriate conditions the nuclear fluid may assume a toroidal shape
\cite{Gam61}. In 1970's the idea of a toroidal nucleus was examined in the
framework of the liquid drop model and shell corrections \cite{Won73,Won78}.

This contribution is devoted to a systematic investigation on the chain of
even-even $Z=120$ isotopes and $N=184$ isotones within the self-consistent
constraint Skyrme-Hartree-Fock-Bogoliubov (Skyrme-HFB) mean-field theory  
at the region of large oblate deformations.

\section{Model and results}
The constrained Skyrme-HFB approach is equivalent to the minimization of the
Skyrme energy density functional $E^{tot}[\boldsymbol{\bar{\rho}}]$ with
respect to the densities and currents under appropriate constraints \cite{Per04}. 
Using the method of Lagrange multipliers we solve an equality-constrained problem (ECP):
\begin{equation}
\left\{
\begin{array}{l}
\displaystyle\min_{\boldsymbol{\bar{\rho}}} E^{tot}[\boldsymbol{\bar{\rho}}]\\
         \mbox{subject to: } \displaystyle\langle \hat{N}_{q} \rangle= N_{q},\quad (q=p,n),\\
\phantom{\mbox{subject to: }}\displaystyle\langle \hat{Q}_{\lambda\mu} \rangle= Q_{\lambda\mu},
\end{array}
\right. 
\label{eq:1}
\end{equation}
where the constraints are defined by the average values $N_{p,n}$ of the proton 
and neutron particle-number operators $\hat{N}_{p,n}$ and by the constrained values 
$Q_{\lambda\mu}$ of the mass-multiple-moment operators $\hat{Q}_{\lambda\mu}$.

The above ECP equations were solved using an augmented Lagrangian method
\cite{Sta10} with the symmetry-unrestricted code HFODD \cite{Sch12}.
In the particle-hole channel the Skyrme SkM* \cite{Bar82} force was applied
and a density-dependent mixed pairing interaction \cite{Sta13} in the
particle-particle channel was used. The code HFODD uses the basis expansion
method utilizing a three-dimensional Cartesian deformed harmonic oscillator basis.
In the present study, we used a basis which consists of states having not
more than $N=26$ quanta in the Cartesian directions, and not more than 1140
states.

\begin{figure}[t]
\begin{center}
\includegraphics[width=0.7\textwidth]{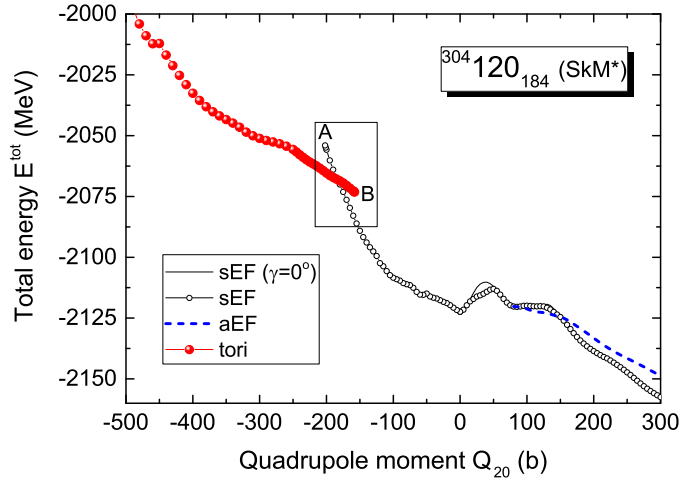}
\caption{\label{fig:1} (Color online.) Total HFB energy of $^{304}120_{184}$
as a function of the quadrupole moment. The open circle symbols and dashed (blue)
line show the symmetric (sEF) and asymmetric (aEF) elongated fission pathways, respectively.
The axially symmetric sEF ($\gamma =0^{\circ}$) fission pathway is marked by a solid thin line.
The nuclear matter density distributions with toroidal shapes appear in the region of large
oblate deformation $Q_{20} \leq -158$ b as (red) solid circles.
}
\end{center}
\end{figure}

As an example, the total HFB energy of SH nucleus $^{304}120_{184}$ as a
function of the quadrupole moment $Q_{20}$ is shown in Fig.~\ref{fig:1}.
In addition to a spherical ground state minimum, one can see two paths
leading to fission on the prolate side: a reflection-symmetric path with the
elongated fission fragments (sEF) (open circles) and a reflection-asymmetric path
with the elongated fission fragments (aEF) (dashed line). On the oblate side,
the self-consistent nuclear density under the $Q_{20}$ constraint changes
from an oblate spheroidal to a biconcave disc shape, as the magnitude of
oblate $Q_{20}$ increases. When the oblate $Q_{20}$ magnitude exceeds 158~b,
there emerges an additional self-consistent toroidal nuclear density solution.

\begin{figure}[t]
\begin{minipage}[b]{0.47\linewidth}
\centering
\includegraphics[width=\linewidth]{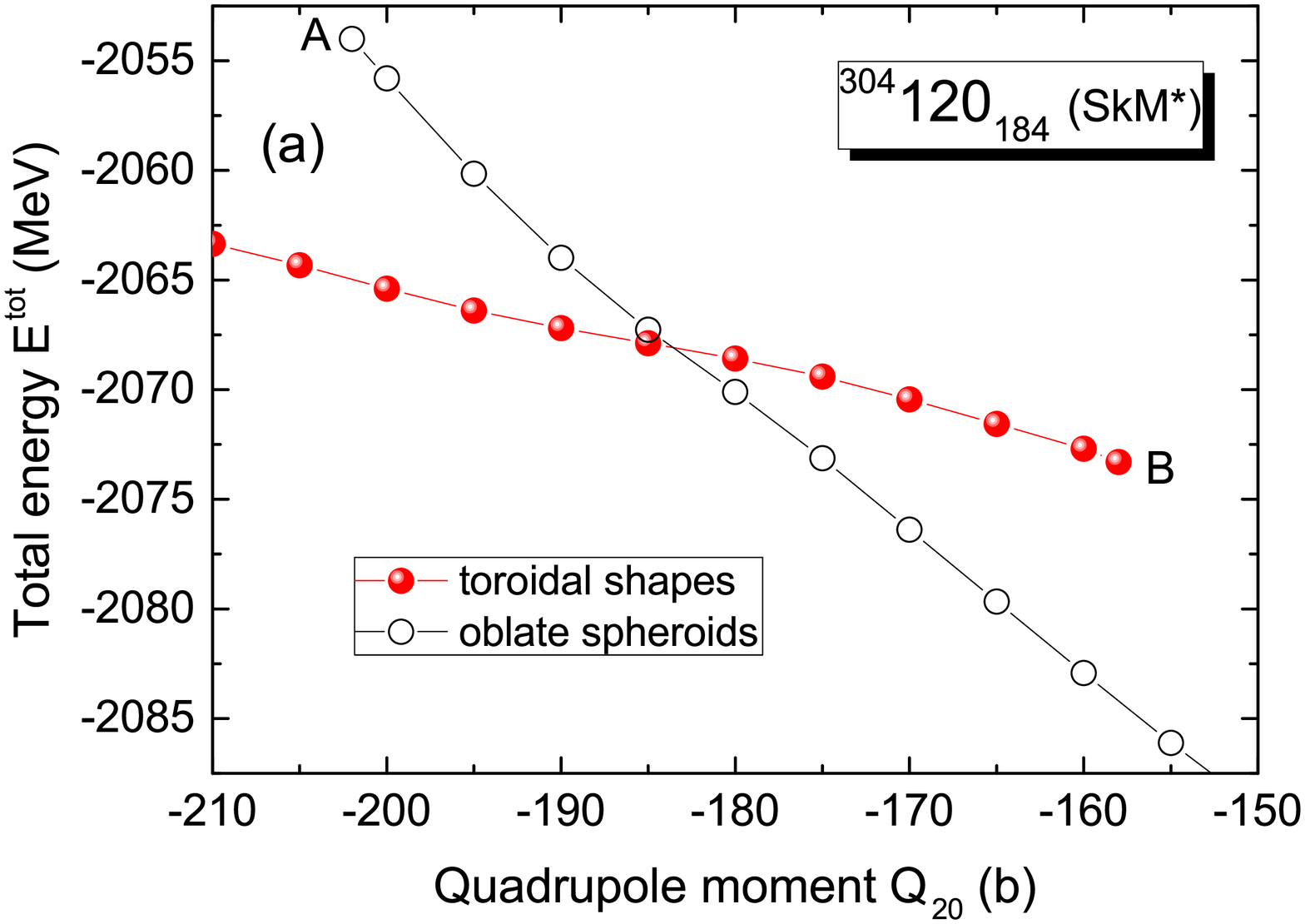}
\end{minipage}%
\hspace*{0.40cm}
\begin{minipage}[b]{0.47\linewidth}
\centering
\includegraphics[width=\linewidth]{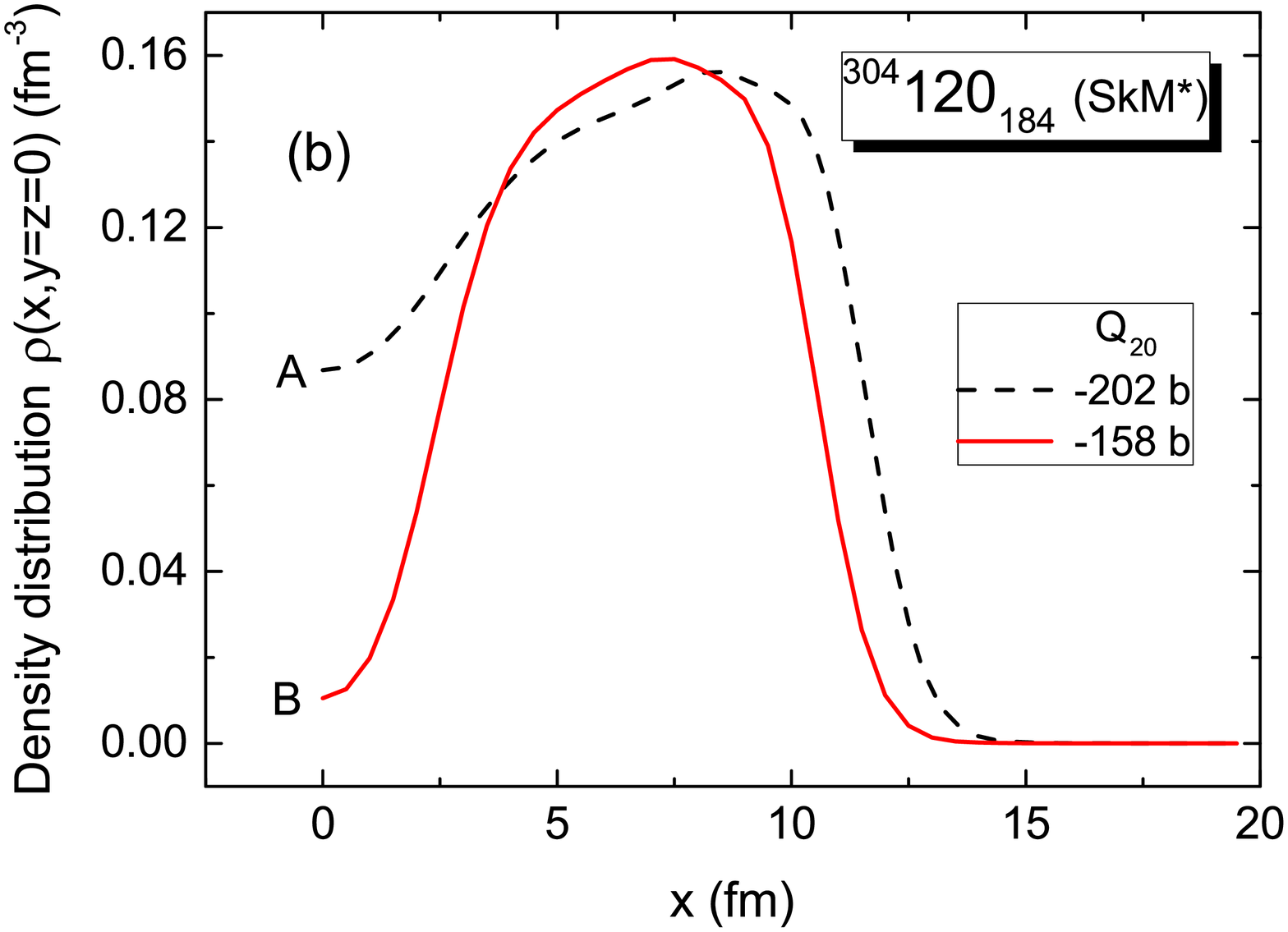}
\end{minipage}
\caption{\label{fig:2ab} (Color online.) Left panel (a), an enlarged view of
Fig.~\ref{fig:1}, the total HFB energy of $^{304}120_{184}$ as a function of the
quadrupole moment between points \textbf{A}-\textbf{B}, where a shape transition
from biconcave discs to toroidal shapes takes place.
Right panel (b), the nuclear density distribution at $Q_{20}({\bf A})=-202$~b
(biconcave disc) and $Q_{20}({\bf B})=-158$~b (torus).
}
\end{figure}

An enlarged view of the transition from the biconcave disc to the toroidal
shape of Fig.~\ref{fig:1} is shown in Fig.~\ref{fig:2ab}(a), where our Skyrme-HFB
calculations give the biconcave disc solutions ending at the point \textbf{A}
(at $Q_{20 }({\bf A})=-202$~b), and another toroidal solutions starting at the
point \textbf{B} (at $Q_{20}({\bf B})=-158$~b).
The nuclear density distributions of $^{304}120_{184}$ calculated at \textbf{A}
and \textbf{B} are depicted in Fig.~\ref{fig:2ab}(b), which indicates that the
nuclear density at the toroidal geometrical center is very small at the starting
point \textbf{B} of the toroidal sequence, but remains to be about 0.08/fm$^3$
at point {\bf A} of the biconcave disc sequence.

\begin{figure}[t]
\begin{center}
\begin{minipage}[b]{0.5\textwidth}
\begin{turn}{-90}
\includegraphics[width=1\linewidth]{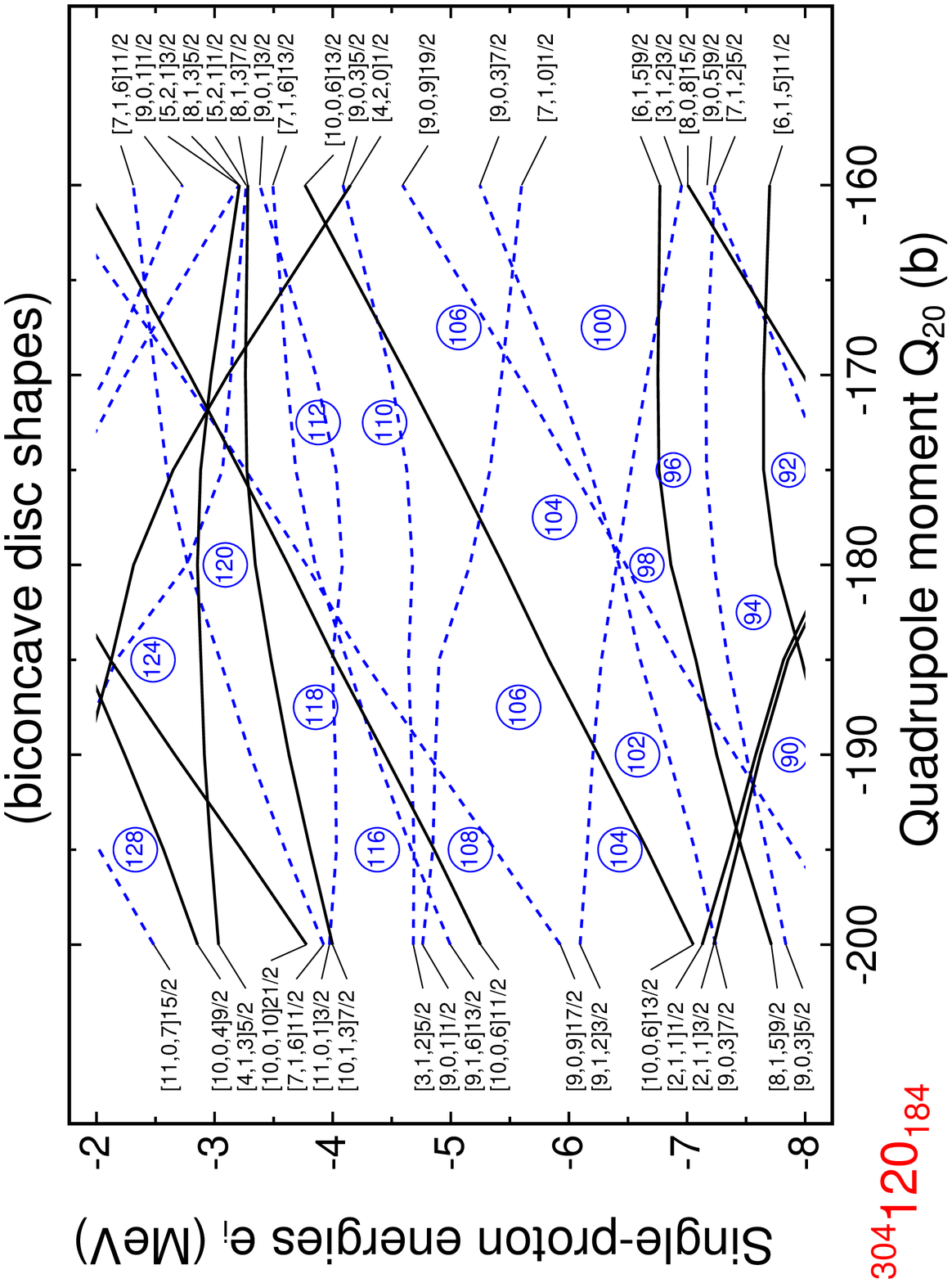}
\end{turn}\\
\vspace*{0.25cm}
\end{minipage}
\begin{minipage}[b]{0.5\textwidth}
\begin{turn}{-90}
\includegraphics[width=1\linewidth]{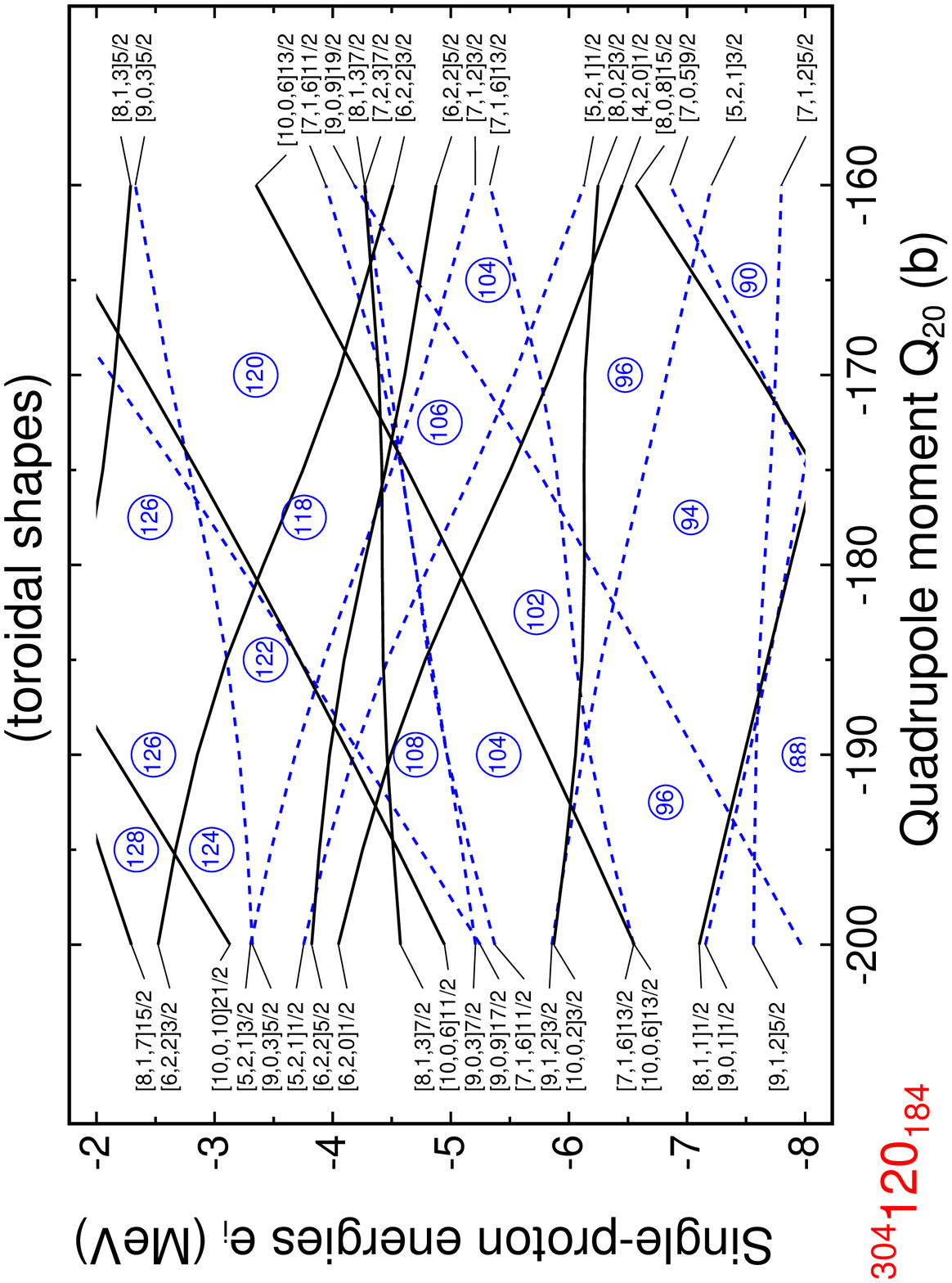}
\end{turn}
\end{minipage}
\caption{\label{fig:3} (Color online.) Proton single-particle levels of
$^{304}120_{184}$ as a function of the quadrupole moment.
The levels with positive parity are drawn with solid lines, while
those with negative parity are drawn with (blue) dashed lines.
The upper panel is for a biconcave disc shape and the lower panel for
a toroidal shape.
}
\end{center}
\end{figure}

\begin{figure}[t]
\begin{center}
\begin{minipage}[b]{0.5\textwidth}
\begin{turn}{-90}
\includegraphics[width=1\linewidth]{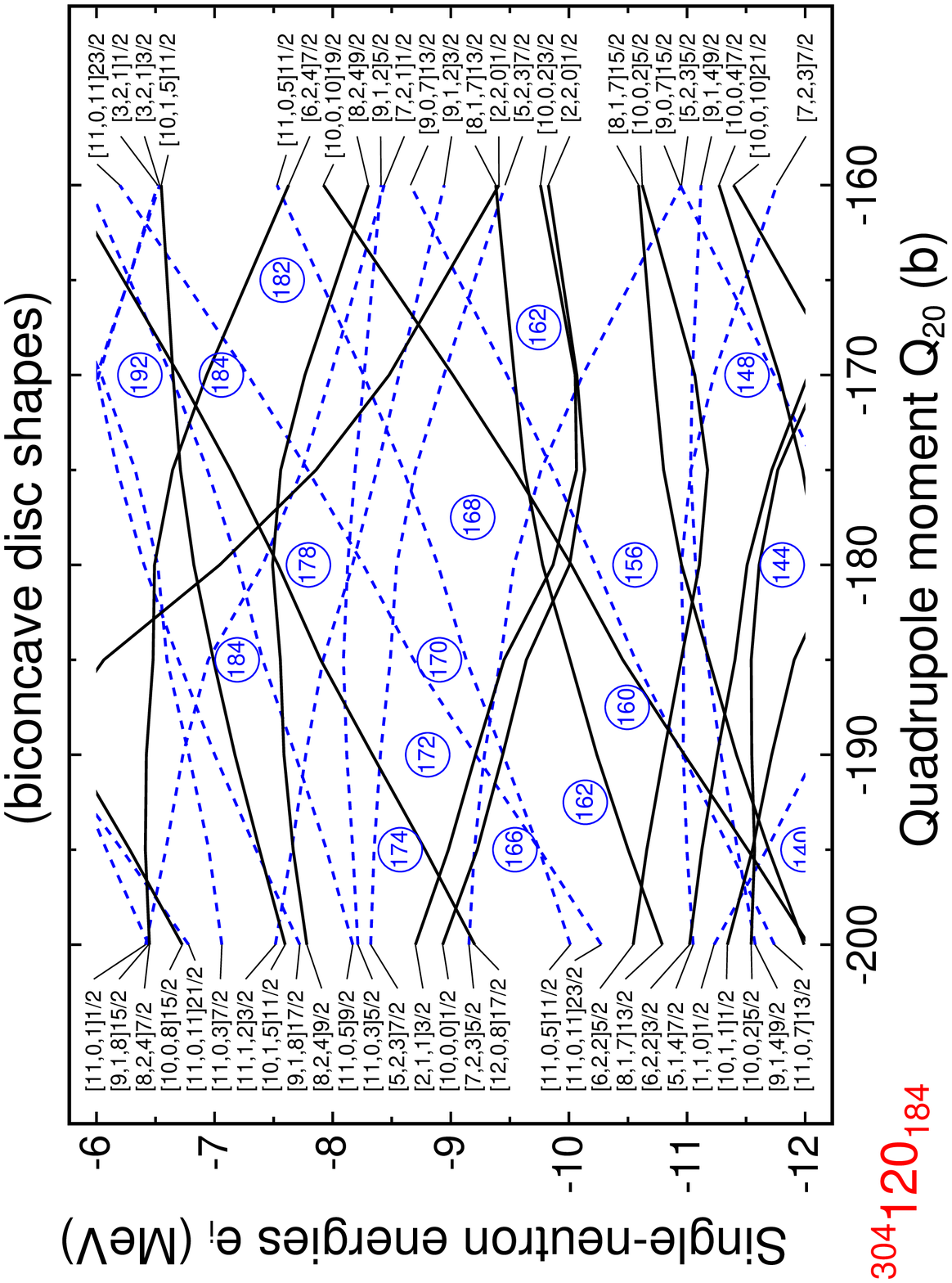}
\end{turn}\\
\vspace*{0.25cm}
\end{minipage}
\begin{minipage}[b]{0.5\textwidth}
\begin{turn}{-90}
\includegraphics[width=1\linewidth]{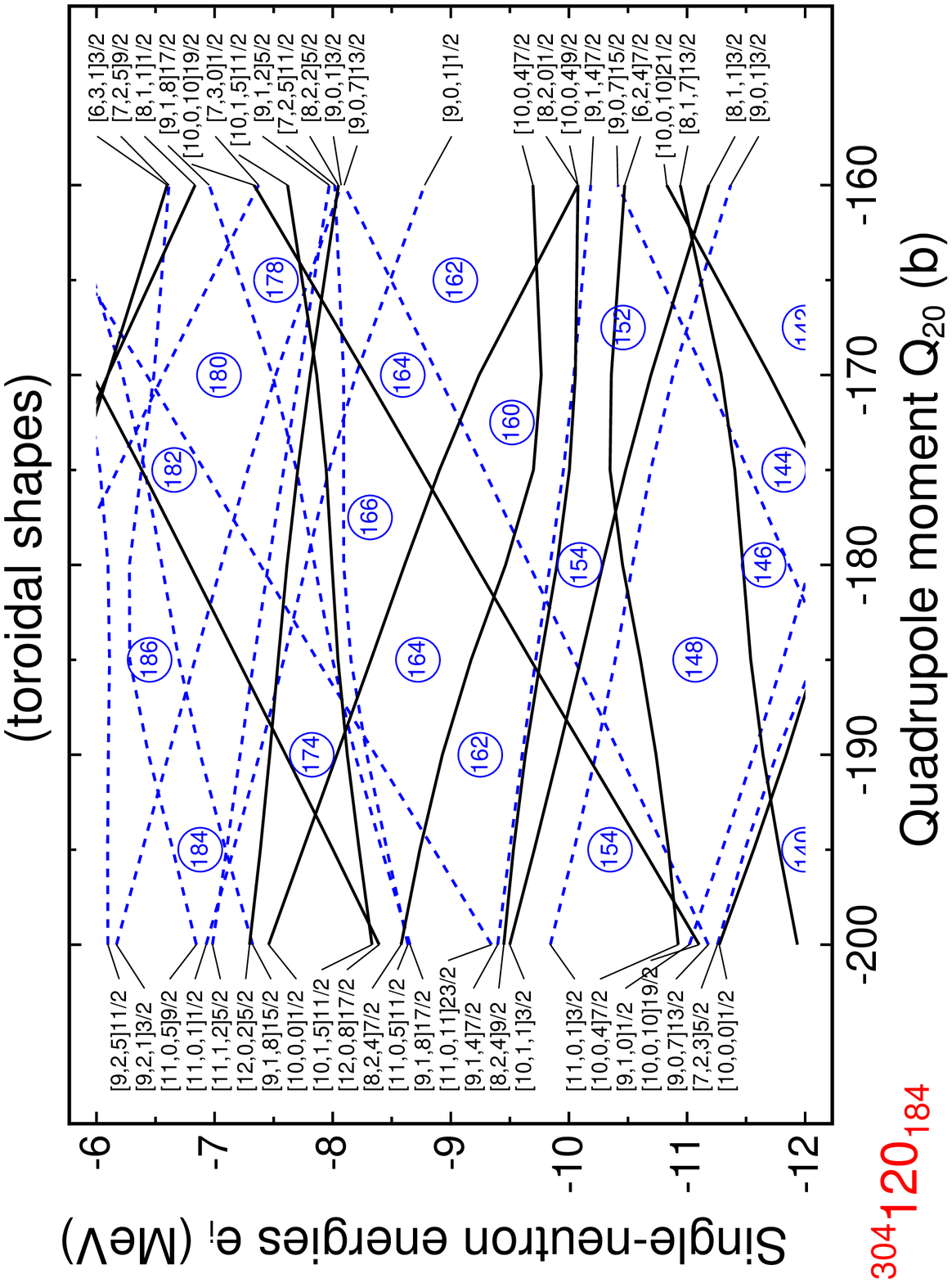}
\end{turn}
\end{minipage}
\caption{\label{fig:4} (Color online.) The same as in Fig.~\ref{fig:3},
but for the neutron single-particle levels.
}
\end{center}
\end{figure}

In the region of quadrupole moment between $Q_{20}({\bf A})$ and $Q_{20}({\bf B})$,
both biconcave disc and toroidal solutions coexist for $^{304}120_{184}$. It is
of interest to examine the single-particle states of these two types of solutions
in this oblate deformation region.
The proton single-particle levels of $^{304}120_{184}$ close to the Fermi energy
as a function of $Q_{20}$ between $Q_{20}({\bf A})$ and $Q_{20}({\bf B})$ for the
biconcave disc and toroidal sequences are shown in the upper and lower panels of
Fig.~\ref{fig:3}, respectively.
Levels with positive parity are drawn as solid curves, while those with negative
parity are drawn as dashed curves. Each single-particle state is labeled by the
Nilsson quantum numbers \break $[N,n_{z},\Lambda]\Omega$ of the dominant component.
Each level is doubly degenerate. The circled numbers denote the occupation numbers.
For the sake of comparison, Fig.~\ref{fig:4} gives the neutron single-particle
levels close to the Fermi energy for the biconcave disc solution in the upper panel
and the toroidal solution in the lower panel.

Even though Figs.~\ref{fig:3} and \ref{fig:4} pertain to the self-consistent
single-particle states for $^{304}120$, we expect that the mean-field potential
depends mostly on nuclear density shape and the quadrupole moment, and
varies only slightly as a function of the atomic number and the neutrons number.
The single-particle state diagrams in Figs.~\ref{fig:3} and \ref{fig:4} can
be approximately applied to an extended region around $^{304}120_{184}$.

We infer from Figs.~\ref{fig:3} and \ref{fig:4} that the densities of proton
and neutron single-particles states are far from being uniform. There are regions
of sparse density of single-particle states which can be identified as single-particle
``shells" associated with enhanced stability \cite{Bra72}. It will be of future
interest to exploit the property of the extra stability of SHN for which the toroidal
proton and neutron shells are located at the same deformation.
The shell effects play an even more significant role when the bulk properties
of the system lead to a nearly flat bulk energy as a function of the deformation,
such as would be expected for systems with $Z\ge 122$ \cite{Sta08}.
The complexity of the single-particle state energy level diagram indicates that the
location of the neutron and proton shells needs to be examined on a case-by-case basis.

\begin{figure}[t]
\begin{minipage}[b]{0.46\linewidth}
\centering
\includegraphics[width=\linewidth]{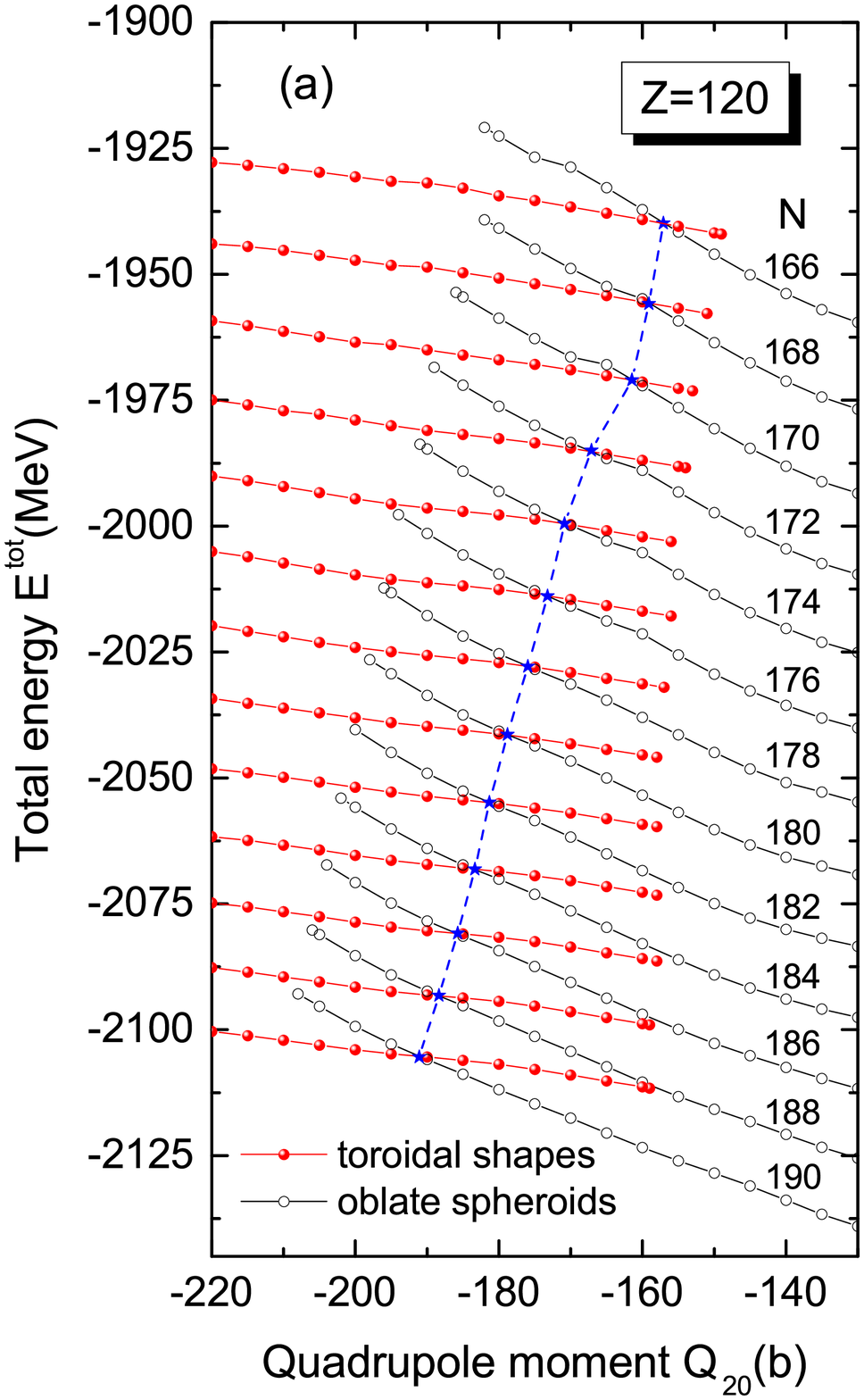}
\end{minipage}%
\hspace*{0.30cm}
\begin{minipage}[b]{0.46\linewidth}
\centering
\includegraphics[width=\linewidth]{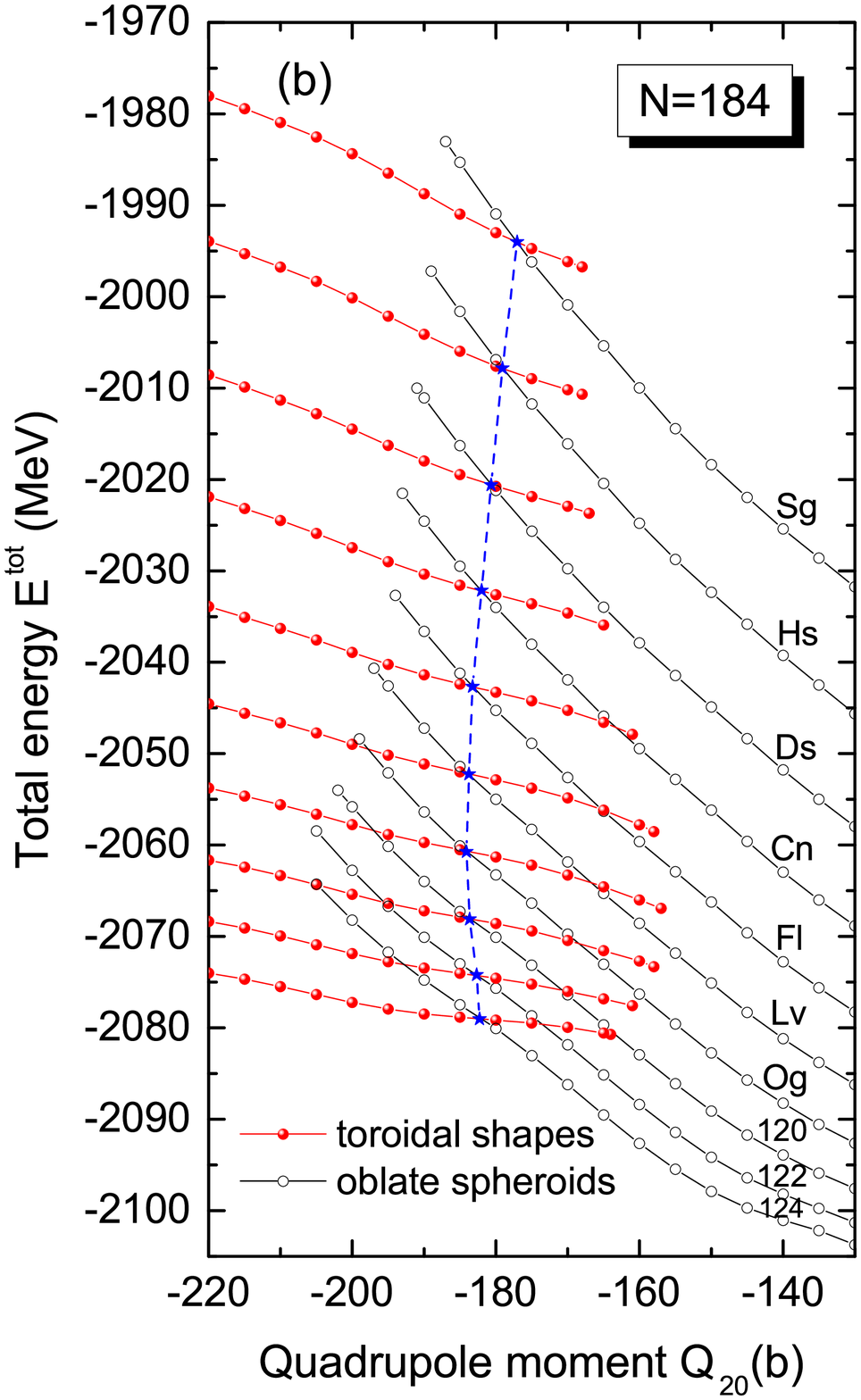}
\end{minipage}%
\caption{\label{fig:7ab} (Color online.) Shape transitions from the biconcave disk
shapes to the toroidal shapes for the even-even isotopes $Z=120$ in panel (a), and for
the even-even isotones $N=184$ in panel (b).
The points of energy crossing of the oblate spheroids and toroidal shapes solutions
are connected by  a(blue) dashed line.
}
\end{figure}

Fig.~\ref{fig:7ab}(a) gives the Skyrme-HFB energies as a function of the quadrupole 
moment of even-even $Z=120$ isotopes with the number of neutrons from 166 to 190, 
in the region of the shape transition from the biconcave disc shape to the toroidal shape.
The toroidal and biconcave disc total energies decreases as a function of increasing 
$Q_{20}$ and do not posses an energy minimum. The slope of the toroidal energy as 
a function of $Q_{20}$ appears to be nearly independent of the neutron number, for 
$Z=120$  isotopes. The biconcave disc energy curve and the toroidal energy curve cross 
each other at an energy crossing point, whose location moves to a more negative $Q_{20}$ 
value as the neutron number increases. Along the lowest energy curve as a function of 
$Q_{20}$ in Fig.~\ref{fig:7ab}(a), there is a sudden shape transition from the biconcave 
disc shape to the toroidal shape at the energy crossing point. The first solution with the 
toroidal shape takes place at $Q_{20}=-150$~b for $N=166$, and at $Q_{20}=-160$~b  for $N=190$.

Similar results, but for the even-even $N=184$ isotones with the number of protons
from 106 to 124 are shown in Fig.~\ref{fig:7ab}(b). For those isotones, one observes
that as the proton number increases, the magnitude of the Coulomb repulsion increases,
and the magnitude of the slope of the toroidal energy curve becomes smaller. The toroidal
energy curve for $Z=124$ is nearly but not completely flat. Further increase in the proton
number may render a toroidal energy equilibrium at a greater oblate deformation. The
energy crossing points of the biconcave disk and toroidal energy curves
occur at $Q_{20}\approx -180$~b for all atomic numbers in Fig.~\ref{fig:7ab}(b).

\section{Conclusions and Discussions}
The Coulomb repulsion from a large number of protons in a SH nucleus has a tendency
to push the nuclear matter outward, making it easier to assume a toroidal shape.
We examine here SHN in the region of $Z=120$ isotopes and $N=184$ isotones. We find
that as the magnitude of the oblate $Q_{20}$ increases along the lowest energy curve,
there is a sudden shape transition from a biconcave disc to a torus.
For $Z=120$ isotopes with $166\le N \le 190$ and for $N=184$ isotones with
$106\le Z \le 124$, the total energy curves lie on a slope, indicating that these nuclei
in a toroidal shape are unstable against returning to the shape of a sphere-like geometry.

Our examination of the single-particle states in this region reveal that the density
of single-particle levels is far from being uniform and single-particle shells are
present at various toroidal deformations. Because the energy curve as a function $Q_{20}$
becomes flatter with increasing $Z$, one expects that by increasing the atomic number
beyond $Z\ge 122$ with possible appropriate toroidal shells, some toroidal figures
of equilibrium may become possible. Future search for toroidal nuclei may focus
attention in this superheavy region in conjunction with possible nuclear shell effects.

The presence of a large angular momentum will facilitate the formation of a toroidal nucleus.
In this regard, it will be useful to examine SHN with non-collective rotations whose
spin along the symmetry axis arises from particle-hole excitations \cite{Boh81}.
Previous investigation reveal a region of toroidal high-spin isomers in the light mass region \cite{Ich12,Ich14,Sta14,Sta15a,Sta15b,Sta16}. A recent investigation shows that
$^{304}120_{184}$ with $I=I_z=81$ and 208$\hbar$ may be toroidal high-spin isomers \cite{Sta17}.

The research was supported in part by the Division of Nuclear Physics,
U.S. Department of Energy under Contract DE-AC05-00OR22725
and the National Science Center, Poland, project no 2016/21/B/ST2/01227.


\end{document}